\documentclass[11pt,a4paper]{article}   
\usepackage{amsmath}
\usepackage{amssymb}
\usepackage{mathrsfs}

\usepackage{rotating}
\usepackage{color}  

\usepackage[round]{natbib}

\usepackage{fancybox}

\newcommand{\real}{\textrm{Re}}

\newcommand{\wrt}{ ~ {\rm d}}

\def\ci{{\mathrm i}}
\renewcommand{\exp}{{\rm e}}
\newcommand{\pn}{\partial_{n}}
\newcommand{\ps}{\partial_{s}}
\newcommand{\px}{\partial_{x}}
\newcommand{\pz}{\partial_{z}}

\newcommand{\lap}{\nabla^{2}}

\newcommand{\ds}{\displaystyle}

\newcommand{\ie}{i.e.\ }

\newcommand{\dint}{\int\hspace{-6pt}\int}
\newcommand{\amp}{A}
\newcommand{\radius}{a}
\newcommand{\fp}{\sigma}
\newcommand{\flex}{F}
\newcommand{\lapsq}{\nabla^{4}}
\newcommand{\sg}{X}
\newcommand{\hv}{Z}
\newcommand{\ph}{\psi}
\newcommand{\dr}{X_{d}}
\newcommand{\vm}{w}
\newcommand{\curv}{\kappa}
\newcommand{\Ahv}{\mathcal{Z}}
\newcommand{\Apt}{\Psi}
\newcommand{\Asg}{\mathcal{X}}
\newcommand{\scatt}{\theta}
\newcommand{\inct}{\hat{\theta}}
\newcommand{\kl}{K}
\newcommand{\cst}{q}

\begin{document}
\title{Water wave transmission by an array of floating disks.}
\author{L.~G.~Bennetts$^{1}$ and T.~D.~Williams$^{2}$
\\
{\footnotesize
$^{1}$School of Mathematical Sciences, University of Adelaide, Adelaide 5005, Australia}
\\
{\footnotesize
$^{2}$Nansen Environment and Remote Sensing Centre, Bergen 5006, Norway}
}
\date{\today}
\maketitle

\begin{abstract}
An experimental validation of theoretical models of transmission of regular water waves by large arrays of floating disks is presented.
The experiments are conducted in a wave basin.
The models are based on combined potential-flow and thin-plate theories, and the assumption of linear motions.
A low-concentration array, in which disks are separated by approximately a disk diameter in equilibrium, 
and a high-concentration array, in which adjacent disks are almost touching in equilibrium, are used for the experiments.
The proportion of incident wave energy transmitted by the disks is presented as a function of wave period, and for different wave amplitudes. 
Results indicate that the models predict wave energy transmission accurately for small-amplitude waves and low-concentration arrays. 
Discrepancies for large-amplitude waves and high-concentration arrays are attributed to wave overwash of the disks and collisions between disks.
Validation of model predictions of rigid-body motions of a solitary disk are also presented.
\end{abstract}


\section{Introduction}


Theoretical models of water wave transmission by large groups of floating thin plates have been developed for over thirty years now.
The models are used to help predict attenuation rates of ocean surface waves in the ice-covered ocean.
However, the models have not yet been thoroughly validated via experimental data.
The lack of validation is conspicuous now that attenuation models are being integrated into large-scale operational forecasting and climate models \citep{Dob&Bid13,Wiletal13a,Wiletal13b,Benetal14a}.

\citet{Squ&Mor80} and \citet{Wadetal88a} used measurements of waves in the ice-covered ocean to
provide evidence that wave energy attenuates approximately exponentially with respect to distance travelled.
Further, the measurements indicate that attenuation rates depend on wave period.
In particular, the ice cover acts as a low-pass filter, 
\ie large period waves maintain energy for the greatest distance into the ice-covered ocean.
 
\citet{Wad73a,Wad86} developed the first theoretical model of wave attenuation due to ice cover. 
The ice cover is modelled as a collection of floating thin plates.
Attenuation results from an accumulation of scattering events.
The scattering events themselves are due to impedance mismatches when a wave travels between a region of open water 
and a region in which a plate covers the water surface.

The model is two-dimensional, with one horizontal dimension and one depth dimension.
Potential-flow theory is used to model the water.
Linear motions are assumed.
Consequently, the boundary conditions that couple motions of the plates and the surrounding water are applied at the 
wetted surfaces of the plates in equilibrium.  

Backscatter is neglected in the model, \ie a single-scattering approximation is applied.
The wave energy transmitted by multiple plates is, thus, the product 
of wave energies transmitted by the individual plates.
The model therefore predicts exponential attenuation,
at a rate proportional to $\ln(T)$, where $T$ is
the proportion of energy transmitted by a single plate.

\citet{Wad73a,Wad86} was only able to approximate the energy transmitted by a single plate  
(and hence the attenuation rate) crudely.
\citet{Mey&Squ94} developed a method to calculate the full-linear solution 
for scattering by a single plate,
and thus, in principle, obtain the attenuation rate predicted by the model of \citet{Wad73a,Wad86}.

\citet{Koh&Mey08a} extended the model of \citet{Wad73a,Wad86} to include multiple wave scatterings, 
\ie multiple reflections and transmissions between plates.
In the multiple-scattering model, unlike the single-scattering model,  
transmission by a single realisation of the plates depends on their configuration.
Disorder in plate locations and properties results in exponential attenuation of wave energy with distance travelled.
\citet{Koh&Mey08a} applied a Monte Carlo algorithm with respect to randomly generated realisations of the plates
to calculate a mean attenuation rate for a given wave period and average properties of the plates.
  
\citet{Ben&Squ12a} developed a similar model to \citet{Koh&Mey08a}, but in which wave phases between plates are considered to be random variables, 
and the attenuation rate, rather than the transmitted energy, is averaged.
They produced an analytic solution for the attenuation rate predicted by the model, in terms of the energy transmitted by an individual plate,
under the assumption that only travelling waves interact between adjacent plates --- the so-called wide-spacing approximation.
The resulting attenuation rate is, somewhat bewilderingly, identical to that predicted by the model of \citet{Wad73a,Wad86}.

The linear potential-flow/thin-plate model has been extended to three dimensions.
The plates are commonly assumed to be disks.
\citet{Mey&Squ96} developed two methods to solve the component model of wave scattering by a single disk.
\citet{Meyetal97} proposed a model of wave energy transport through an array of identical disks, based on a single-scattering assumption.

\citet{Pet&Mey09a}, \citet{Ben&Squ09b} and \citet{Benetal10} developed a three-dimensional model that includes multiple wave scattering.
However, the model relies on an artificial periodicity in one horizontal dimension.
A less restrictive multiple-scattering model is yet to be developed.
The role of multiple scatterings in three dimensions is, hence, unclear.

\citet{Koh&Mey08a}, \citet{Benetal10} and \citet{Ben&Squ12b} compared model predictions of attenuation rates
against attenuation rates extracted from field data, which were reported by \citet{Squ&Mor80} and \citet{Wadetal88a}. 
The data were recorded in 1979, and were, until recently, the most complete data sets available on wave attenuation in the ice-covered ocean.
The comparisons indicate broadly reasonable model/data agreement for mid-range wave periods \citep[6\,s to 15\,s estimated by][]{Koh&Mey08a}.
However, the comparisons are far from comprehensive in terms of the properties of the ice cover and wave field.
Notably, the experimental data lacks information on incident wave amplitudes.

\citet{Kohetal14} recently reported attenuation rates extracted from field measurements made in 2012.
The measurements indicate that attenuation rates vary significantly between small- and large-amplitude waves.

Here, results of a laboratory experimental campaign to study transmission of regular water waves by large arrays of floating disks are reported.
The results are compared to predictions given by two- and three-dimensional linear potential-flow/thin-plate models.
The investigation represents, to the authors' knowledge, the first experimental validation of the models, 
as models of water wave transmission by an array of thin floating disks, as opposed to models of wave attenuation in the ice-covered ocean.

The investigation also contains the first validation of rigid-body motions of a solitary disk in response to wave forcing, as predicted by the models.
\citet{Monetal13a,Monetal13b} provided the only previous related validation.
However, \citet{Monetal13a,Monetal13b} applied artificial restraints to the disk to match the model assumptions, as part of the experimental set-up.
In particular, a rod was used to restrict surge motions, 
and a barrier was attached  to the edge of the disk to prevent wave overwash.

The model-data comparisons presented here indicate that linear potential-flow/thin-plate theory predicts wave transmission accurately 
for small incident wave amplitudes and low disk concentrations.
For large incident wave amplitudes and high concentrations the wave energy transmitted in the experiments
is significantly less that that predicted by the models.
The discrepancies are attributed to the unmodelled dissipative processes of wave overwash of the disks and collisions between disks.


\section{Experimental design}

\begin{figure}
 \centering
 \includegraphics[width=0.7\textwidth]{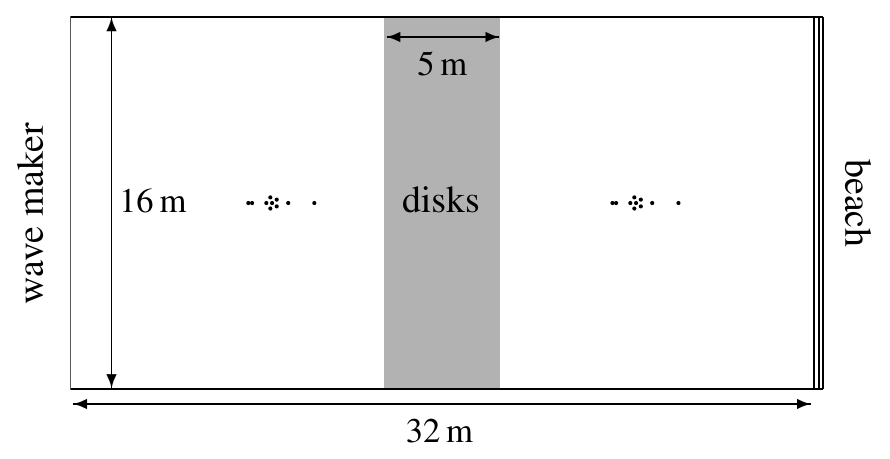}
 \caption{\label{fig:BGOschem}Schematic plan view of wave basin. Dots indicate wave probe locations. Ambient fluid depth is 3.1\,m.}
\end{figure}   

\begin{figure}
 \centering
 \includegraphics[width=\textwidth]{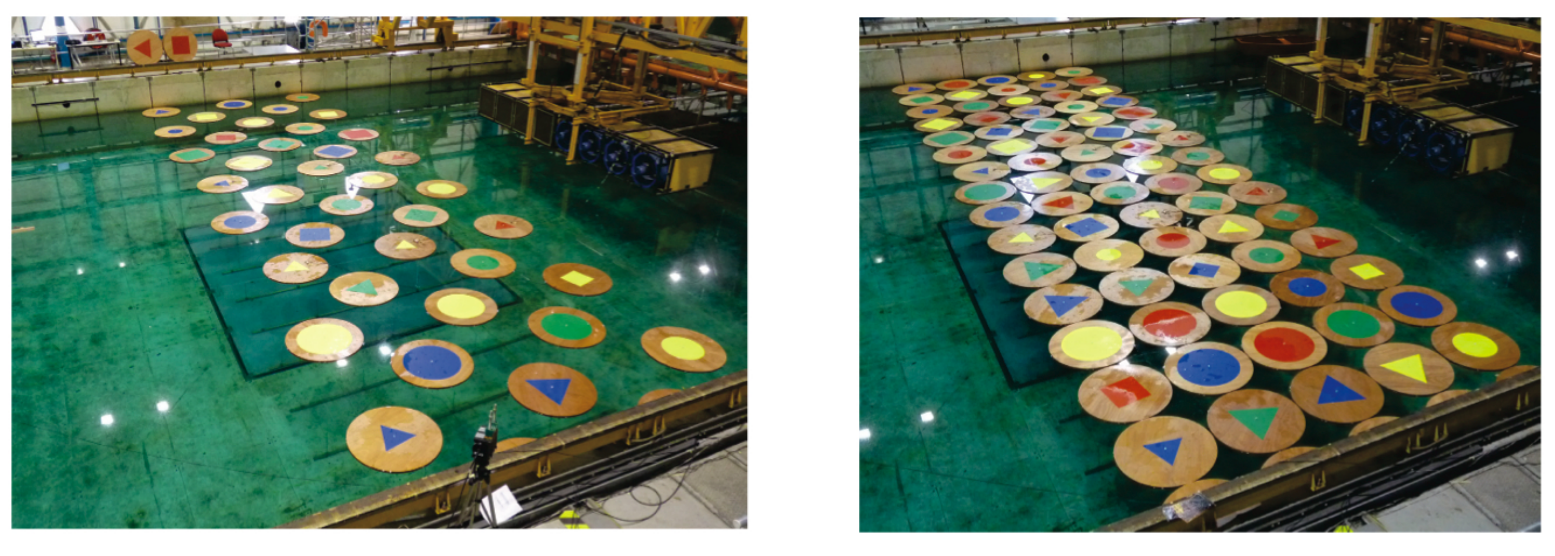}
 \caption{\label{fig:disks}Photos of disk arrays: low concentration (left-hand panel) and high concentration (right).}
\end{figure}

\begin{table}
\centering
\includegraphics[width=\textwidth]{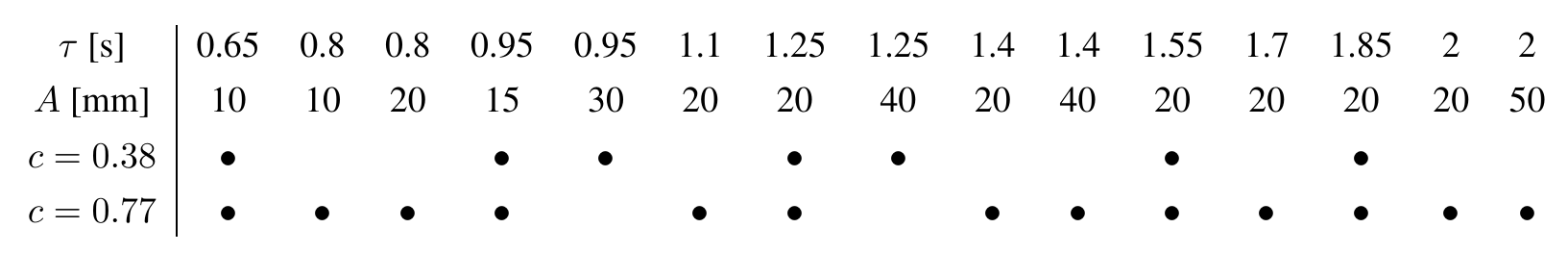}
\caption{\label{tab:tests}Summary of tests conducted, with respect to wave period, $\tau$, and amplitude, $a$.}
\end{table}

The experimental campaign was conducted using the Basin de G\'{e}nie Oceanique \textsc{first} wave basin facility, located at Oceanide, La Seyne sur Mer, France.
The wave basin is 32\,m long, bounded at one end by a unidirectional wave maker and at the opposite end by a static beach. 
The basin is 16\,m wide, bounded by reflective walls. 
The ambient fluid depth in the basin was set at $h=3.1$\,m.
Figure~\ref{fig:BGOschem} shows a schematic plan view of the wave basin.

Wooden disks were installed in the wave basin.
The disks have radius $\radius=0.495$\,m and thickness $D=33$\,mm.
The disks have mass of approximately 14\,kg, and a draught of $d=18$\,mm.
The Young's modulus of the wood used for the disks was measured as $E=4$\,GPa, using a cantilevered beam.
The disks are, therefore, approximately rigid.

Each disk was loosely moored to the basin floor.
The mooring system consisted of a series of three springs and a steel cable.
The natural period of the moored disks was measured as $\tau_{m}=12.5$\,s, 
which is at least an order of magnitude greater than the wave periods used.
The mooring therefore permitted the disks to respond naturally in rigid-body motions to wave forcing, 
but returned the disks to the same initial locations for each test.

A low- and a high-concentration array of the disks were considered.
Eighty disks were used for the high-concentration array. 
The centres of the disks formed a 1\,m by 1\,m square lattice with five rows and sixteen columns, 
\ie adjacent disks almost touching in equilibrium.
The equilibrium concentration of the disks on the water surface was approximately $c=0.77$.
The low-concentration array was formed by removing alternate disks from the high-concentration array.
The equilibrium concentration in this case was approximately $c=0.38$.
Figure~\ref{fig:disks} shows photos of the two arrays.

Regular waves of a prescribed target period, $\tau$, and target amplitude, $\amp$, were generated by the wave maker.
Two groups of probes were used to measure the wave field before and after the disks.
Figure~\ref{fig:BGOschem} indicates the location of the probes.
The groups consist of (i) a line of probes in the direction of the incident wave, 
and (ii) a pentagon of probes, to analyse the directional spectrum of the waves (not considered here, see \S\,\ref{sec:multidisks}).
The probes have an accuracy of at least 1\,mm, and a sampling frequency 250\,Hz.

Target wave periods used for the tests were in the range 0.65\,s to 2\,s.
Target wave amplitudes of 10\,mm, 15\,mm and 20\,mm were used for wave periods 0.65\,s to 0.8\,s, 0.95\,s and 1.1\,s to 2\,s, 
respectively.
Larger wave amplitudes were also used for selected wave periods.
For all period/amplitude combinations the wave steepness was less than 0.05 to avoid wave breaking.
Table~\ref{tab:tests} provides a summary of the tests conducted.
A subset of the tests were repeated.

Accelerometers were attached to six of the disks.
Four of the accelerometers were triaxial and two were biaxial, in the plane of the disk surface.
The six chosen disks comprised two disks from the front row, two from the middle row and two from the back row.
The disks were in the off-centre columns, and hence formed two lines of three disks.

Tests were also conducted for a single disk.
The target periods and amplitudes used for the single-disk tests were identical to those used for the low-concentration array tests.
The Krypton motion tracking system was used to record the six rigid-body motions of the disk during the tests. 
Three light-emitting diodes (\textsc{led}s) were attached to the disk. 
The \textsc{led}s were monitored by a camera, mounted on a platform approximately two metres away from the disk.
The Krypton system constructs time series of translational and rotational motions of the disk from the coordinates of the \textsc{led}s.
The translations have a maximum error of 1\,mm, and the rotations 0.1 degree. 


\section{Theoretical model}


\subsection{Preliminaries}

\begin{table}
\centering
\includegraphics[width=0.45\textwidth]{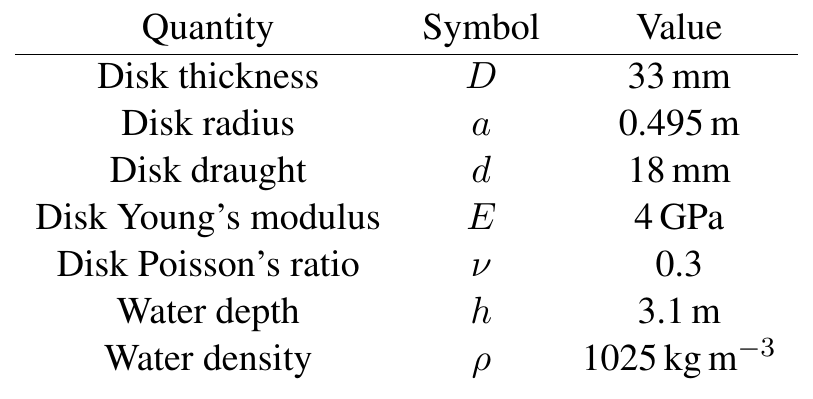}
\caption{\label{tab:params}Disk and water properties used in model calculations.}
\end{table}

Consider a plane wave incident on a group of floating disks, for a water domain of finite depth and laterally unbounded.
The aim is to calculate the proportion of incident wave energy transmitted by the disks.

A Cartesian coordinate system $(x,y,z)$ is used to define positions in the water, where $(x,y)$ are the horizontal coordinates and 
$z$ is the vertical coordinate.
The vertical coordinate points upwards and has its origin set to coincide with the undisturbed free surface of the water.
Under wave motion, the location of the free surface is defined as $z=\zeta(x,y,t)$,  where $t$ denotes time.
A flat impermeable floor bounds the lower surface of the water. 
The location of the floor is $z=-h$. 

The water is assumed to be homogeneous, incompressible, inviscid and in irrotational motion.
It follows that the velocity field can be defined as the gradient of a velocity potential, $\Phi(x,y,z,t)$. 
Time-harmonic motions, of prescribed angular frequency $\omega=2\pi/\tau$,  are assumed.
The velocity potential is therefore expressed as $\Phi(x,y,z,t)=\real\{(g/\ci \omega)\phi(x,y,z)\exp^{-\ci\omega t}\}$.
Here $\phi(x,y,z)$ is a (reduced) velocity potential and $g\approx 9.81$\,m\,s$^{-2}$ is acceleration due to gravity.
Similarly, the free surface is expressed as $\zeta(x,y,t)=\real\{\eta(x,y)\exp^{-\ci\omega t}\}$, where $\eta(x,y)$ is the (reduced) surface displacement.

It is assumed that the incident wave steepness is sufficiently small that 
motions can be modelled accurately by linear theory.
In the absence of the disks, the velocity potential satisfies Laplace's equation throughout the undisturbed water domain, \ie
\begin{subequations}\label{eqns:water}
\begin{equation}
\lap\phi
=0
\qquad
(-h<z<0).
\end{equation}
The velocity potential and surface displacement are coupled at the equilibrium surface by the dynamic and kinematic conditions
\begin{equation}\label{eqn:surfconds}
\phi
=
\eta
\quad
\text{and}
\quad
\pz\phi
=
\fp
\eta
\qquad
(z=0),
\end{equation}
\end{subequations}
respectively, where $\fp=\omega^{2}/g$ is a frequency parameter.

Kirchhoff-Love thin-plate theory is used to model the disks.
The motions of the disks are therefore obtained from the displacements of their lower surfaces.
Consider an individual disk 
and assume, without loss of generality, that the disk is centred on the origin of the horizontal coordinate system when in equilibrium.
The displacement of the disk is denoted $z=-d+\real\{\eta(x,y)\exp^{-\ci\omega t}\}$. 
The interval in which Laplace's equation holds is adjusted in the disk-covered water region to account for draught, \ie
\begin{subequations}\label{eqns:plate}
\begin{equation}
\lap\phi
=0
\qquad
(-h<z<-d).
\end{equation}
The conditions that couple the velocity potential and displacement function are, similarly, recast on the lower surface of the disk.
The dynamic condition is also extended to incorporate disk cover. 
The surface equations in the disk-covered region are thus
\begin{equation}
\phi
=
\eta
\quad
\text{and}
\quad
(1-\fp d)\eta + \flex \lapsq \eta
=
\fp \phi
\quad
(z=-d),
\end{equation}
\end{subequations}
where 
\begin{equation}
\flex=\frac{E D^3}{12\rho g(1-\nu^2)}
,
\end{equation} 
is a scaled flexural rigidity of the disk, $\rho\approx1025$\,kg\,m$^{-3}$ is water density, and $\nu\approx 0.3$ is Poisson's ratio. 
(All disk and water properties used in the model are summarised in table~\ref{tab:params}.)
Free-edge conditions are applied at the disk boundary, which ensure bending moments are shear stresses vanish.
The free-edge conditions are
\begin{equation}\label{eqn:freeedge}
\lap\eta
-
(1-\nu)
\left(
\ps^{2}\eta
+
\curv
\pn
\eta
\right)
=0
\quad
\text{and}
\quad
\pn\lap\eta
+
(1-\nu)
\ps
\pn
\ps
\eta
=0
,
\end{equation}
for $(x,y)\in\Gamma=\{(x,y):x^{2}+y^{2}=\radius^{2}\}$, 
where $\curv$ is curvature of the boundary, and $\pn$ and $\ps$ are the normal and tangential derivatives, respectively.

Let the incident wave originate at $x\to-\infty$, and propagate in the positive $x$-direction.
The corresponding velocity potential is $\phi_{I}=\amp\exp^{\ci k x}\vm(z)$, where $\vm(z)=\cosh k(z+h) / \cosh(kh)$.
The wavenumber $k$ is the positive real root of the dispersion relation
\begin{equation}\nonumber
k\tanh(kh)=\fp.
\end{equation}

Hydrodynamic pressure on the submerged portions of the disks, caused by wave motion,
forces the disks to surge back and forth. 
The location of the centre of the disk, which has its equilibrium centre set to coincide with the origin of the horizontal coordinate system, is hence
$(x,y)=(\real\{ \sg\exp^{-\ci\omega t}\},0)$. 
The amplitude of surge motion, $\sg$, is coupled to the velocity potential via the kinematic condition
\begin{equation}\label{eqn:surge}
\px\phi
=
\fp\sg
\qquad
(-d<z<0),
\end{equation}
for $(x,y)\in \Gamma$,
and the equation of motion
\begin{equation}\label{eqn:EOM}
-\fp d \pi \radius^{2} \sg
=
\int_{\Gamma}
\int_{-d}^{0}
[\phi]_{(x,y)\in\Gamma}
\wrt z
\wrt s.
\end{equation}

The velocity potential and displacement function are obtained by solving equations (\ref{eqns:water}a--b) in the open water regions, 
and equations (\ref{eqns:plate}a--b) in the disk-covered water regions, 
augmented by the bed condition
\begin{equation}
\pz\phi=0
\qquad
(z=-h),
\end{equation}
in both regions.
Conditions (\ref{eqn:freeedge}) and (\ref{eqn:surge}) are imposed at the disk boundaries, 
and an equation of motion of the form (\ref{eqn:EOM}) is invoked for each disk to determine the surge amplitudes.
Appropriate radiation conditions must also be applied to ensure scattered waves travel away from the disks.


\subsection{Two-dimensional model}

A simplified, two-dimensional model is derived by neglecting motions in the $y$-direction, 
and using a geometry that is representative of a cross section in the $(x,z)$ plane. 
Each individual disk then partially reflects and partially transmits incident waves.
Reflection and transmission coefficients for a single `disk'  can be obtained using the method of \citet{Ben&Chu11}, for example.

Reflection and transmission properties of individual disks can be combined to calculate reflection and transmission properties of the group of disks
\citep{Ben&Squ12a}.
Full interaction theory accounts for multiple reflections and transmissions between disks.

Two alternative models are considered here.
First, a model in which backscatter is neglected, \ie a single-scattering model \citep{Wad73a,Wad86}.
Second, a model in which multiple wave reflections and transmissions are included \citep{Ben&Squ12a}. 
In the latter, wave phases between disks are considered to be random variables, and
the energy transmitted by a group of disks is calculated as an average over all possible phases between disks \citep{Ben&Squ12a}.
Remarkably, both models predict the identical energy transmitted by the group of disks.
The transmitted energy proportion is
\begin{equation}
|T|^{cL/\radius},
\end{equation}
where $T$ is the transmission coefficient
for a single disk of radius $\radius=0.495$\,m, draught $d=18$\,mm and Young's modulus $E=4$\,GPa.
The quantity 
$L=5$\,m is the distance travelled by the wave.
Note that the two-dimensional model predicts that the logarithm of the transmitted energy is proportional to the concentration of the disks, $c$.


\subsection{Three-dimensional model}

In the three-dimensional problem, each disk scatterers wave energy across the directional spectrum.
The scattered wave field for an individual disk can be calculated using the method of \citet{Mon12}, for example. 
Let the proportion of incident wave energy scattered at angle $\scatt$, with respect to the positive $x$-axis, 
be denoted $\kl(\scatt)$.

\citet{Meyetal97} and \citet{Mey&Mas06} developed the Boltzmann model for 
wave energy transport through a region of floating disks.
The model extends the scattering model for an individual disk to a large group of disks,
using a single-scattering approximation \citep{How60}.

The Boltzmann model considers the directional spectrum of average wave energy 
with respect to random realisations of disk locations, denoted $S$.
Here, the average wave energy is assumed to be independent of location across the basin width,
\ie no wall effects, 
and to represent the steady regime, \ie independent of time, thus $S=S(x,\inct)$. 

Setting the disk-covered region to occupy the interval $0<x<L$, 
the Boltzmann model in the present setting is 
\begin{equation}\label{eqn:Boltz}
\cos(\scatt)
\px
S(x,\scatt)
=
\ds
\cst
S(x,\scatt)
+
\frac{c}{\pi \radius^{2}}
\int_{-\pi}^{\pi}
\kl(\scatt-\inct)
S(x,\inct)
\wrt\inct
\quad
(0<x<L).
\end{equation}
The left-hand side of equation~(\ref{eqn:Boltz}) provides steady advection of the wave spectrum.
The right-hand side provides scattering of wave energy.
The quantity $\cst$ 
represents the proportion of incident wave energy not contained in the scattered wave field, \ie
\begin{equation}
\cst =
\frac{c}{\pi \radius^{2}}
\int_{-\pi}^{\pi}
\kl(\inct)
\wrt\inct
.
\end{equation}
Incident wave energy from the wave maker side of the disks is specified by the boundary conditions
\begin{equation}
S(0,\scatt)
=
\amp^2
\delta(\scatt)
\quad
(\vert\scatt\vert<\pi/2)
\quad
\text{and}
\quad
S(L,\scatt)
=
0
\quad
(\vert\scatt\vert>\pi/2),
\end{equation}
where $\delta$ denotes the Delta function.
(Note that incident wave energy has been normalised to the incident amplitude squared.)
The wave energy spectrum, $S$, is calculated using a discretisation of the angular coordinate, $\scatt$,
and applying a generalised spectral method to the resulting system of ordinary differential equations
\citep{Meyetal97,Mey&Mas06}.


\section{Results}\label{sec:results}


\subsection{Data processing}\label{sec:stft}

\begin{figure}
 \centering
 \includegraphics[width=\textwidth]{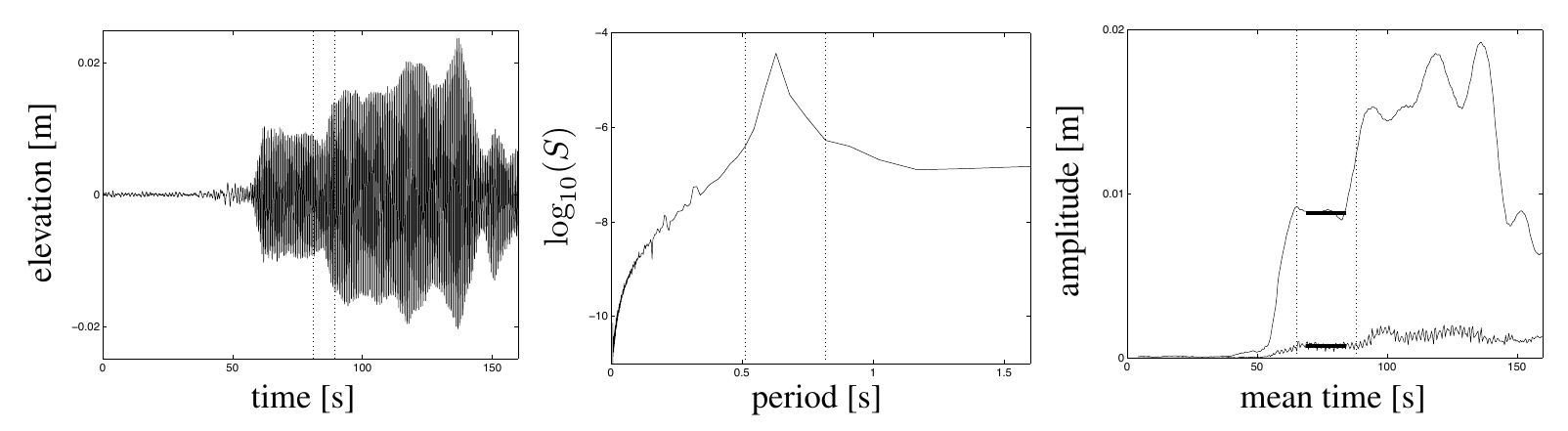}
 \caption{\label{fig:spec}Example conversion of time series to representative amplitude in  steady-state interval. 
 Left-hand panel shows raw surface elevation time series from probe closest to wave maker, for test with target wave period $0.65$\,s
 and high-concentration array.
 Dotted vertical lines indicate a time window.
 Middle panel shows the energy density for the indicated window.
 Dotted vertical lines indicate bins used to calculate the amplitude at the target period.
 Right-hand panel shows the amplitude at the target period and half the target period (smaller) as functions of mean time. 
 Dotted vertical lines indicate the steady-state interval. 
 Bars indicate the extracted amplitude and chosen interval. }
\end{figure}

A short-time Fourier transform (\textsc{stft}) method is used to covert raw time series of surface elevations, 
and disk translations and rotations 
into spectral representations of amplitudes in time, \ie spectograms.
Representative amplitudes for chosen wave periods and time intervals are extracted from the spectrograms.

For a given time series, amplitudes are calculated using the following algorithm.
\begin{enumerate}
\item
The series is windowed. 
A target window width is specified.
The actual window width used is that closest to the target width, which contains a power of two samples. 
\item
The inverse fast Fourier transform is applied to the series in each window.
The (generalised) energy density spectrum, $S(\tau,\hat{t})$, where $\hat{t}$ denotes mean time for windows,  
is calculated as twice the square of the modulus of the transformed series.
\item
The energy at the target period
is calculated as the sum of the energy in the bin containing that period, 
and energies contained in up to three bins either side of that bin.
The corresponding amplitude is the square-root of two times the energy.
\item
A time interval of interest is defined. 
Typically, a steady-state interval is selected.
The steady-state interval begins after transients in the leading waves have passed, and ends before the signal is contaminated 
by reflections from geometrical boundaries.
A representative amplitude for the interval of interest is calculated as the mean of the amplitudes in the interval.
\end{enumerate}

Figure~\ref{fig:spec} shows an example of conversion from time series to amplitudes.
The left-hand panel shows a single window on the raw time series.
Window width is chosen to balance time localisation (narrow windows) and spectral resolution (wide windows).
The target width in the example shown is ten target wave periods.
(Target widths are three wave periods for the largest target wave periods.)

The middle panel shows the energy spectrum in the example window.
A peak around the target wave period, $\tau=0.65$\,s, is evident.
Spectral leakage is also evident.

The right-hand panel shows amplitudes of the target wave period and half the target period, \ie the first and second harmonics, as functions of mean time.
The steady-state interval is also indicated.
In the case considered, the lower bound of the interval is the time at which the incident waves reach the probe, and the upper bound is the time at which the first waves reflected by the disks reach the probe.
The interval is calculated analytically via the group velocity.
The amplitudes vary near to the boundaries of the interval,
which is partially due to smearing produced by the \textsc{stft} method.
(The chosen example shows particularly strong variation.) 
Representative amplitudes are therefore extracted from a smaller interval.
In this case, the interval is the middle two-thirds of the full steady-state interval.


\subsection{Single disk: response amplitude operators}\label{sec:raos}

\begin{figure}
 \centering
 \includegraphics[width=\textwidth]{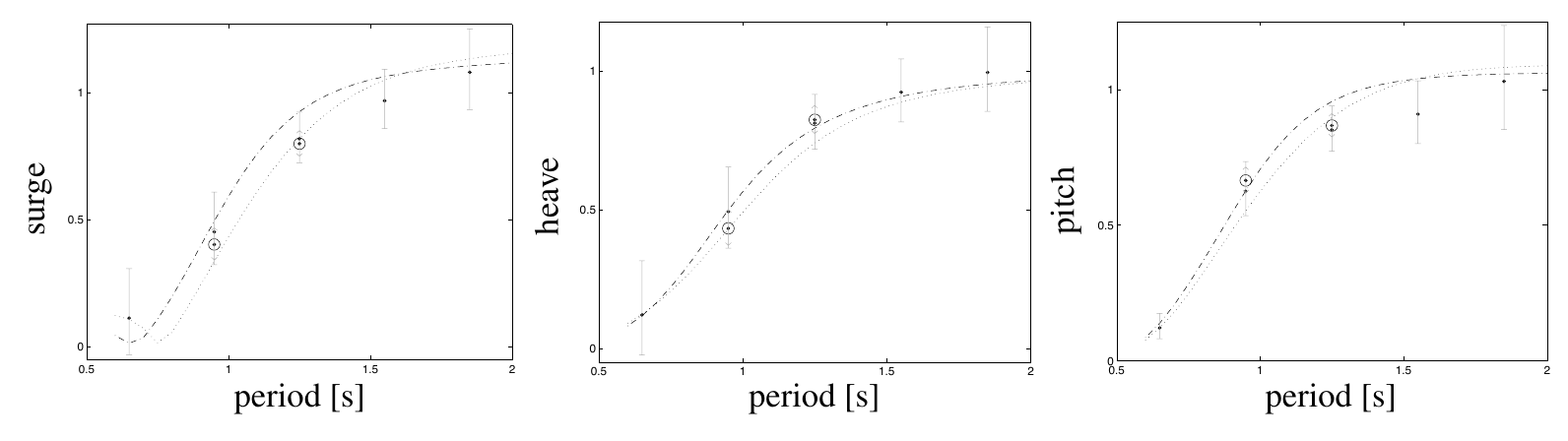}
 \caption{\label{fig:RAOs}\textsc{Rao}s for a single disk. 
 Bullets denote experimental data. Encircled bullets denote data from large-amplitude tests.
 Grey bars denote error bounds from measuring devices.
 Curves denote results of theoretical models. Dotted curves are results from the two-dimensional model and chained curves are results from the three-dimensional model. }
\end{figure}


\begin{figure}
 \centering
 \includegraphics[width=\textwidth]{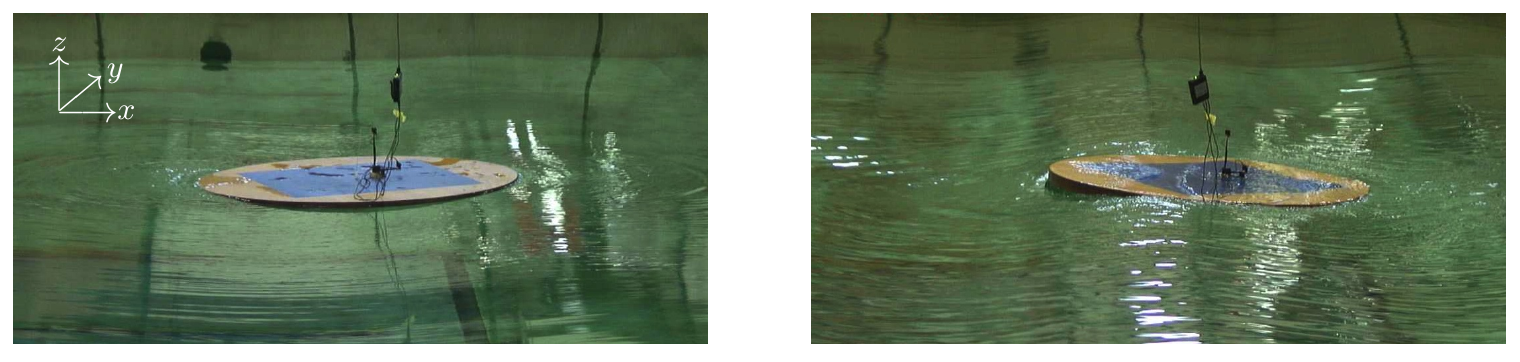}
 \caption{\label{fig:overwash} Example of onset of overwash with increased wave amplitude.
 Photos are of single disk experiments for wave period $0.95$\,s and amplitudes $15$\,mm (left-hand panel) and 30\,mm (right).}
\end{figure}

Let time series of disk translations in the $x$- and $z$-directions be denoted $\Asg(t)$ and $\Ahv(t)$, respectively, 
and the angle of rotation of the surface about the $y$-axis be denoted $\Apt(t)$.
In the steady-state interval the translational and rotational motions are 
\begin{equation}
\Asg\approx \real\{\dr\exp^{-2\pi\ci  t/\tau_{m}}+\sg\exp^{-2\pi\ci  t/\tau}\}, 
\quad
\Ahv\approx \real\{\hv\exp^{-2\pi\ci  t/\tau}\}
\quad
\text{and}
\quad 
\Apt\approx \real\{\ph\exp^{-2\pi\ci  t/\tau}\}.
\end{equation}
The quantity $\dr$ is the amplitude of the disk motion related to the mooring system, which is excited by drift forces.
Drift is not analysed here.
The quantities $\sg$ and $\hv$  are, respectively, the amplitudes of surge and heave motions of the disk. 
The quantity $\ph$ is the amplitude of angle of rotation about the $y$-axis, and is the arctangent of the amplitude of pitch motion of the disk.
The amplitudes $\sg$, $\hv$ and $\ph$ are extracted from time series for $\Asg$, $\Ahv$ and $\Apt$ using the algorithm described in \S\,\ref{sec:stft}.

The amplitude of surge motion is calculated explicitly in the theoretical models.
Amplitudes of heave and pitch motions are obtained from the displacement function, $\eta$, via
\begin{equation}
\hv
=
\frac{1}{\pi \radius^{2}}
\dint_{\Omega}
\eta(x,y)
\wrt x\hspace{-3pt}\wrt y
\quad
\text{and}
\quad
\tan(\ph)
=
\frac{1}{4\pi \radius^{4}}
\dint_{\Omega}
x \eta(x,y)
\wrt x\hspace{-3pt}\wrt y,
\end{equation}
for the three-dimensional model,
where $\Omega$ is the lower surface of the disk \citep{Monetal13b}.
Analogous expressions exist for the two-dimensional model.

The response amplitude operators (\textsc{rao}s) associated to surge, heave and pitch motions are 
$\vert\sg\vert / \amp$, $\vert\hv\vert / \amp$ and $\vert\tan(\ph)\vert / k\amp$, respectively. 
The amplitude of the incident wave, $\amp$, for a given test, is calculated as the mean amplitude in the steady state intervals 
of the time series given by the ten probes on the wave maker side of the disk, prior to arrival of reflected waves. 

Figure~\ref{fig:RAOs} shows a comparison of the  \textsc{rao}s calculated from the experimental data and the theoretical models.
Error bounds for the experimental data are obtained by adding/subtracting the maximum errors in the measuring devices from the calculated translations, 
rotations and incident wave amplitudes. 

Generally good agreement is found between the models and data.
Agreement is marginally better for the three-dimensional model than the two-dimensional model.
The two models are similar over the chosen range of wave periods.
The two-dimensional model tends to suppress motions in comparison to the three-dimensional model.
However, surge and pitch \textsc{rao}s in the two-dimensional model exceed those of the three-dimensional model for large wave periods. 

Tests using wave periods  $0.95$\,s and 1.25\,s were conducted for two wave amplitudes.
In both cases, the larger amplitude was twice the smaller amplitude.
For tests using wave period $1.25$\,s, the two amplitudes produce almost identical \textsc{rao}s .
For tests using wave period $0.95$\,s, although the two amplitudes produce similar \textsc{rao}s, differences are notable. 
The larger amplitude reduces the translational, surge and heave \textsc{rao}s and increases the rotational, pitch \textsc{rao}.

Wave overwash of the disk is a visible effect of increasing amplitude.
Overwash refers to the wave running over the top of the disk.
It is not included in the models.
Overwash provides an explanation of the change in the \textsc{rao}s for the large-amplitude test using a wave period 0.95\,s. 
Translational motions are suppressed by the extra load of the overwashed fluid on the disk surface.
Conversely, pitch motions are enhanced, as rotations force front and back ends of the disk (with respect to the incident wave) to be submerged alternately.
Overwashed water originates from the submergences.

Figure~\ref{fig:overwash} shows photos of the disk in the tests using wave period $0.95$\,s.
Full overwash is evident for the large-amplitude test.
For the instant captured by the photo, the angle of rotation of the disk is maximal with respect to the negative $x$-axis.
Overwashed water is therefore generated at the back end of the disk in this phase of its motion.
Only a small quantity of water is visible on the disk surface for the small-amplitude test.
Overwash also occurs during the large-amplitude test using wave period $1.25$\,s, and the test using wave period 0.65\,s.  
However, in both cases, the overwash is weaker, in terms of depth and disk coverage, than the large-amplitude test using wave period $0.95$\,s. 


\subsection{Multiple disks}\label{sec:multidisks}

\begin{figure}
 \centering
 \includegraphics[width=\textwidth]{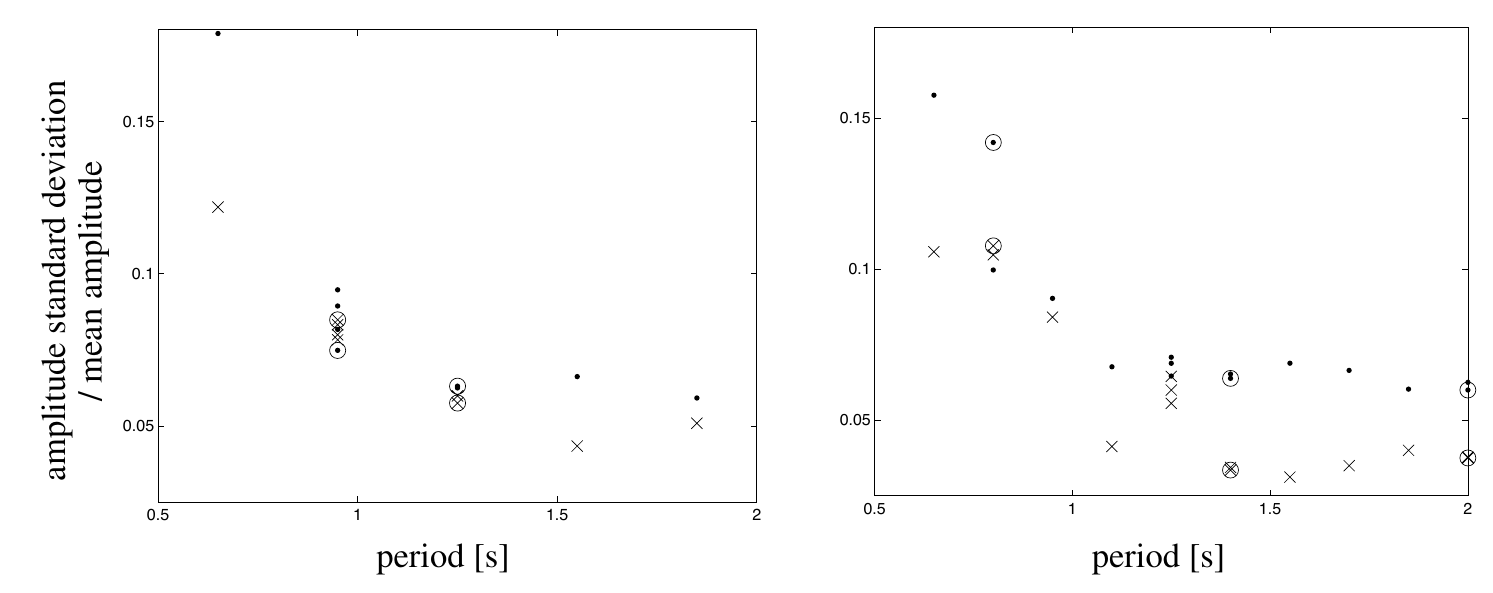}
\caption{\label{fig:Tstd} Standard deviations in wave amplitudes scaled with respect to mean amplitudes for low-concentration array tests (left-hand panel) and high-concentration array tests (right).
 Dots denote transmitted waves.
 Crosses denote incident waves. Circles denote large-amplitude tests. }
\end{figure}

\begin{figure}
 \centering
 \includegraphics[width=\textwidth]{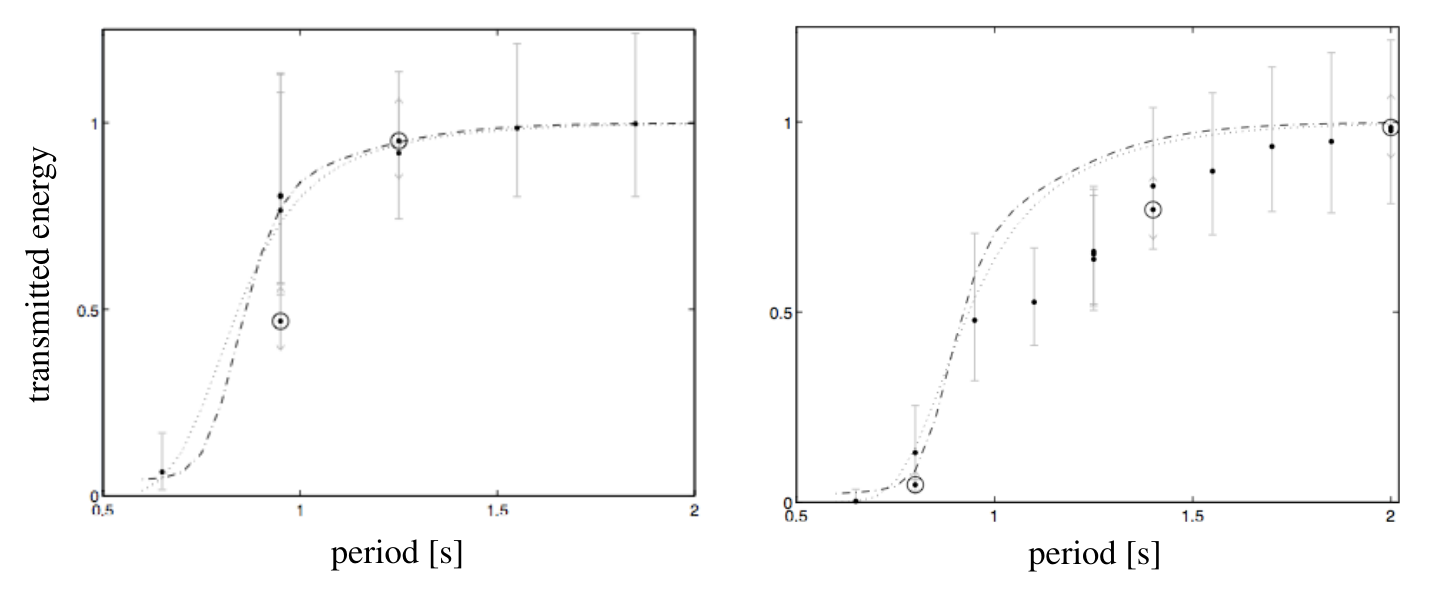}
\caption{\label{fig:Trans} Transmitted energy proportion for low-concentration array tests (left-hand panel) and high-concentration array tests (right). 
 Curves and symbols as in figure~\ref{fig:RAOs}.}
\end{figure}

\begin{figure}
 \centering
 \includegraphics[width=\textwidth]{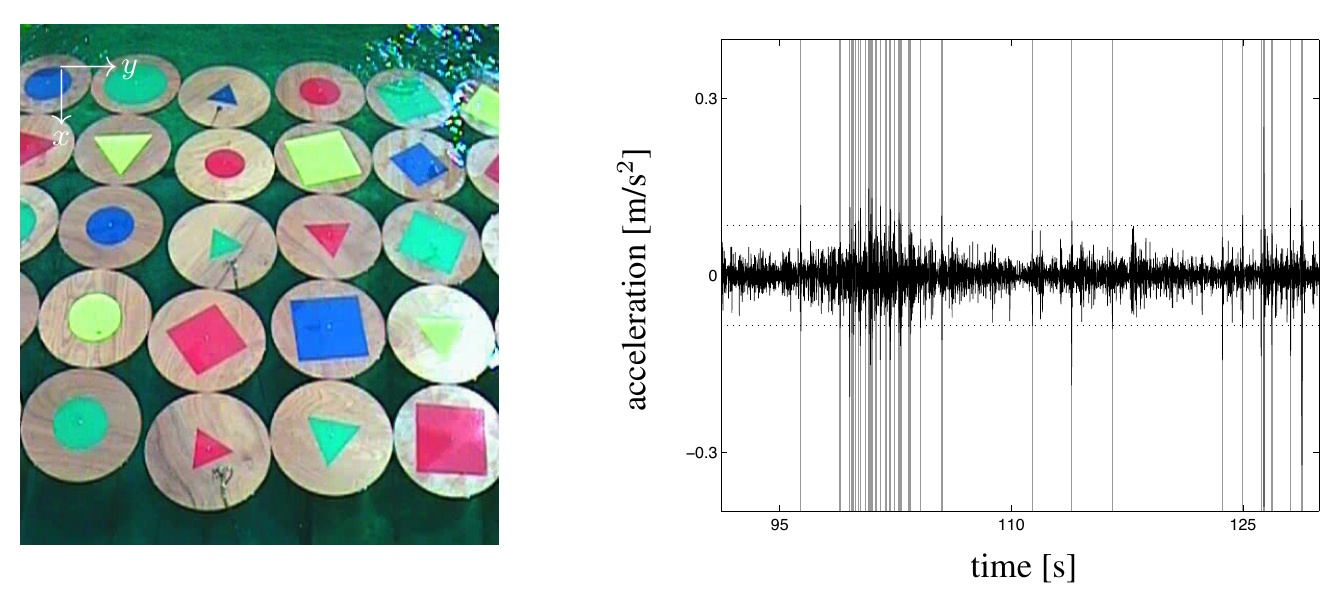}
 \caption{\label{fig:test06} Photo at time 100\,s of subset of disks in a high-concentration array test, 
 centred around disks with accelerometers in right-of-centre column of full array, 
 using wave period 0.65\,s and amplitude 10\,mm  (left-hand panel),
 and high-pass filtered acceleration time series for disk in middle row (right).
 Thresholds of collisions (black-dotted, horizontal lines) and locations of collisions (grey, vertical lines) are overlaid on acceleration time series.}
\end{figure}

\begin{figure}
 \centering
 \includegraphics[width=\textwidth]{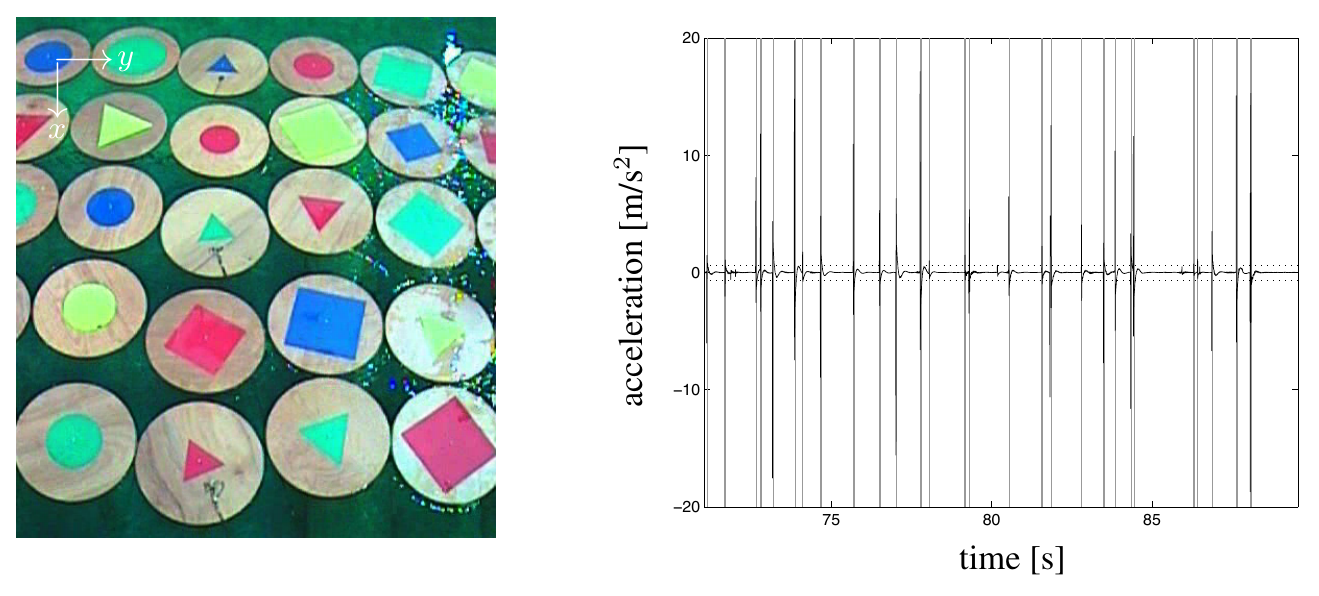}
 \caption{\label{fig:test11} As in figure~\ref{fig:test06} but for test using wave period 1.25\,s and amplitude 20\,mm.
 Photo taken at time 74\,s.  }
\end{figure}

\begin{figure}
 \centering
 \includegraphics[width=\textwidth]{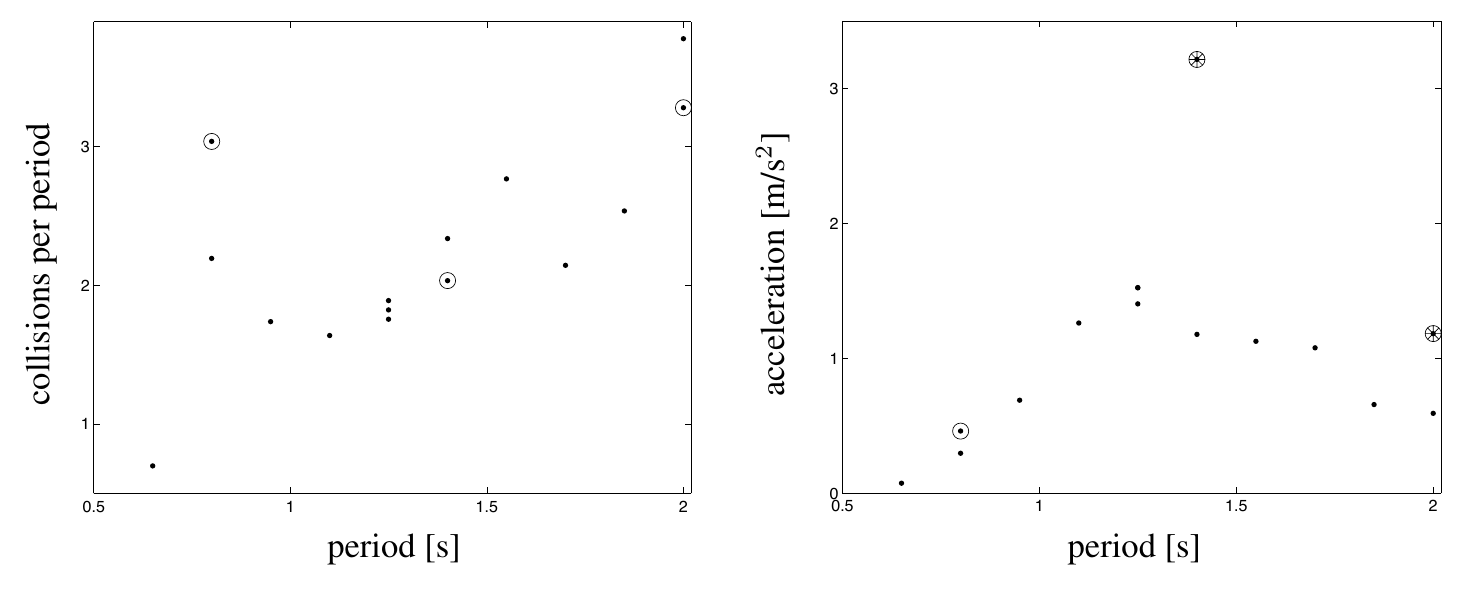}
 \caption{\label{fig:CollStats} Mean number of collisions experienced by disks with accelerometers in middle row of high-concentration array per target wave period (left-hand panel),
 and corresponding mean magnitude of accelerations (right). Asterisks denotes tests in which one or more collisions saturates the signal. 
 }
\end{figure}

\begin{figure}
 \centering
 \includegraphics[width=\textwidth]{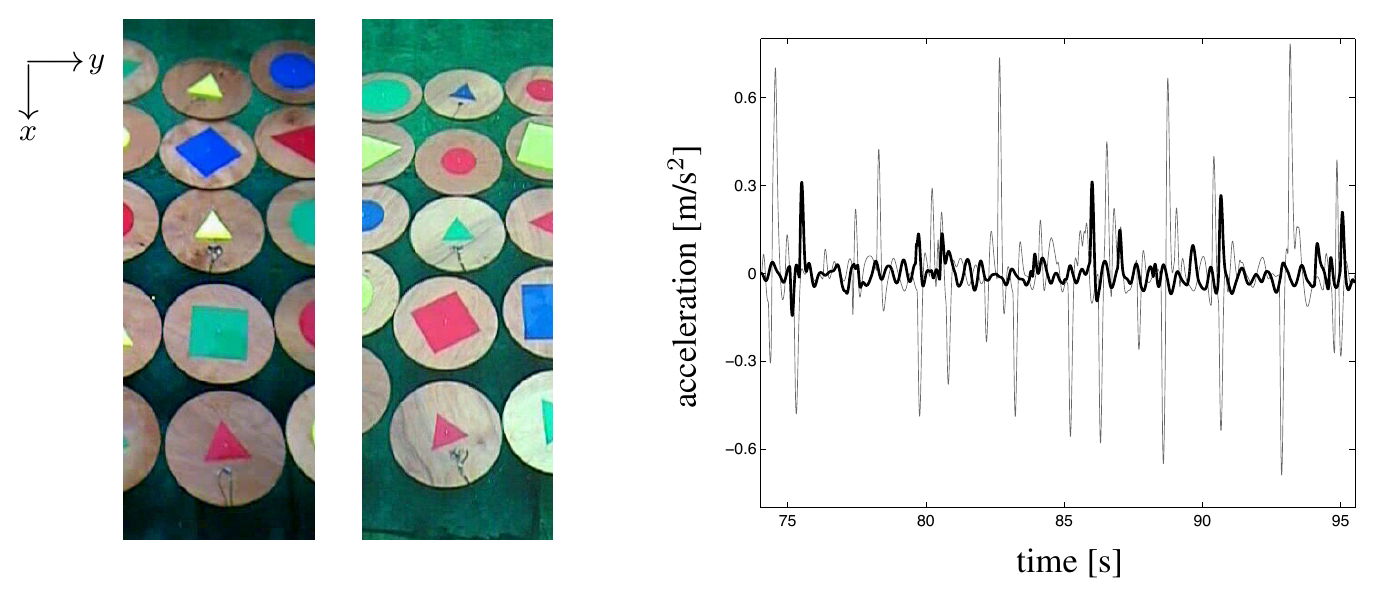}
 \caption{\label{fig:Rafts} Example of rafting event in test using wave period 1.1\,s and amplitude 20mm. 
 Left-hand photo shows disks in left-of-centre column, with three top disks rafted, 
 and right-hand photo shows right-of-centre column, with no disks rafted.
 Right-hand panel shows low-pass filter response of accelerometers in middle of left column (black curve) and right column (grey). }
\end{figure}

A transmitted wave amplitude, $\amp_{T}$, is calculated from the experimental data as the mean of the amplitudes in the steady state intervals 
of the time series given by the ten probes on the beach side of the disks.
Figure~\ref{fig:Tstd} shows the standard deviations in transmitted amplitudes over the array of probes, scaled with respect to the corresponding mean amplitude.
Scaled standard deviations are also shown for the incident waves, \ie amplitudes from the probes on the wave maker side of the array.

The transmitted wave fields were visibly dominated by a plane wave in the incident wave direction for all tests, except for those using the smallest wave period, $\tau=0.65$\,s.
Geometrically decaying circular wave fields emanating from the back row of disks were, however, also visible for tests using wave periods less than or equal to 1.25\,s.
Plane wave dominance of the transmitted wave fields at the probes in the majority of the tests is confirmed by the scaled standard deviations.
Standard deviations are less than 10\% of their corresponding mean amplitudes for all tests, except those using the smallest wave period,  $\tau=0.65$\,s, 
and the large-amplitude test using wave period 0.8\,s for the high-concentration array.
Further, standard deviation to mean ratios are almost identical to their incident wave counterparts for the low-concentration tests, 
with the exception of the test using wave period 0.65\,s. 
For tests using the high-concentration array,  standard deviation to mean ratios are appreciably larger for the transmitted wave fields than the incident wave fields, 
by up to a factor of approximately 2.2.
Differences are particularly notable for tests using wave periods greater than or equal to 1.4\,s.

Standard deviations of transmitted amplitudes for tests in which plane wave dominance was not visible were less than 0.5\,mm, 
\ie beneath the resolution of the wave probes.
Analysis of the directional spectrum of the transmitted wave fields is, therefore, not conducted here. 

The proportion of energy transmitted by the disks is calculated as $(\amp_{T}/\amp)^2$.
The two-dimensional models predict the transmitted energy  $(\amp_{T}/\amp)^2 = |T|^{cL/\radius}$.
For the Boltzmann model, the transmitted energy is calculated as the energy contained in the transmitted wave spectrum in a small band around the incident wave angle, \ie
\begin{equation}
\amp_{T}^{2}
=
\int_{-\scatt_{0}}^{\scatt_{0}}
S(L,\scatt)
\wrt
\scatt
.
\end{equation}
The band width parameter is $\scatt_{0}=\pi/100$ in the results presented .

Figure~\ref{fig:Trans} shows a comparison of the proportion of energy transmitted by the disks, 
calculated from the experimental data and the theoretical models.
The incident amplitude, $A$, is calculated from the experimental data, in this case, as the mean  steady-state amplitude given by the ten probes on the beach side of the disks
in tests conducted without the array of disks.
The method is chosen to account for small attenuation of waves simply in travelling from the wave maker to the probes.
Error bounds for the experimental data are obtained as for the \textsc{rao}s.

The two- and three-dimensional models give almost identical predictions of the transmitted energy in the chosen interval of wave periods.
Both predict monotonic increase in transmission as wave period increases.
Small differences between the predictions are evident for wave periods approximately less than 1.25\,s for the low-concentration array
and less than 1.8\,s for the high-concentration array.
The transmitted energies interleave one another in the intervals that they differ. 
Consequently, there is no clear characteristic behavioural difference for transmission between the two- and three-dimensional models.


Excellent agreement is found between the models and data for the low-concentration array of disks.
The transmitted energy for the large-amplitude test using  wave period  0.95\,s is the only exception to the excellent agreement.
This is a consequence of significant amplitude dependence for tests using wave period 0.95\,s. 
The increase in amplitude from 15\,mm to 30\,mm causes the transmitted energy to reduce by a factor of approximately 0.6.
The effect of increasing amplitude found  here for multiple disks is far more pronounced than for the \textsc{rao}s of a single disk seen in \S\,\ref{sec:raos}, as may be anticipated.
The amplitude dependence is again attributed to overwash.
For the large-amplitude test, all disks were overwashed, despite attenuation of the wave amplitude with distance travelled through the array.

In comparison, the increase in amplitude from 20\,mm to 40\,mm for tests using wave period 1.25\,s produces a very small increase in the transmitted energy, 
which is well within the bounds of uncertainty.
All disks were overwashed in the large-amplitude test.
However, as in the corresponding single-disk test, the overwash was weak.
The results therefore indicate that weak overwash has a negligible effect on wave energy transmission.


The models and data both predict that the high-concentration array transmits less wave energy than the low-concentration array.
However, agreement between the models and data is significantly worse for the high-concentration array.
The models and data do agree for the test using the largest wave period, $\tau= 2$\,s, although, in this case, the incident wave is fully transmitted.
Note that the incident wave is fully transmitted in both the small- and large-amplitude tests. 
Good agreement is also found for tests using wave periods less than 0.95\,s, where strong attenuation occurs.

Amplitude dependence is evident for the tests using wave period 0.8\,s.
Here, increasing the amplitude from 10\,mm to 20\,mm 
results in the proportion of wave energy transmitted being reduced by a factor of approximately 0.35.
However, the impact of the reduction is debatable due to strong attenuation of the small-amplitude wave.
Moreover, it is difficult to argue whether the models agree more closely with the small- or large-amplitude test.

Amplitude dependence is also evident for tests using wave period 1.4\,s.
However, the dependence is weak.
Increasing the amplitude from 20\,mm to 40\,mm results in 
wave energy transmission being reduced by a factor of approximately 0.93.

The models and data differ most for tests using wave periods in the mid-range interval 0.95\,s to 1.85\,s. 
The experimental data indicates less energy is transmitted than the models. 
Small differences exist for tests using wave periods 1.7\,s and 1.85\,s. 
It is notable that the models predict full transmission for these periods, 
whereas the experiments indicate 1.5\% to 2.3\% energy attenuation.
Differences increase for tests using wave periods 1.4\,s (small amplitude) and 1.55\,s, 
with the models overestimating transmission by a factor of approximately 1.1.
Differences are greatest for periods 1.1\,s, 1.25\,s and 1.4\,s (large amplitude), 
with the models overestimating transmission by factors 1.5, 1.4 and 1.2, respectively.


Discrepancies between models and data for mid-range wave periods
cannot be attributed simply to extrapolation of small discrepancies for tests using the low-concentration array to tests using the high-concentration array.
Instead, it is likely that wave energy is attenuated due to collisions between disks. 
Adjacent disks are separated by approximately 10\,mm in the high-concentration tests, 
which makes the disks susceptible to collisions under wave forcing.
Collisions did not occur in the low-concentration tests.
Attenuation due to collisions is not included in the models.

The strength and frequency of collisions during the tests are quantified here using data provided by the accelerometers mounted on the two disks in the middle row of the array. 
Only the data from the accelerometers initially in $x$-direction are used, \ie the direction of the incident waves.
Collisions are detected using the following algorithm.
\begin{enumerate}
\item
The steady state intervals of the acceleration time series are identified, as for the wave probes.
\item
A third-order Butterworth filter is used to separate the high- and low-frequency components of the series.
\item
Thresholds of two times the significant accelerations of the low-pass filters, \ie four standard deviations, are set.
\item
Collisions are identified as accelerations in the high-pass filters with magnitudes greater than or equal to the thresholds.
(Collisions occurring repeatedly for up to ten consecutive time steps are regarded as the same collision, \ie collisions are allowed a duration of 0.04\,s.) 
\end{enumerate}
Figures~\ref{fig:test06} and \ref{fig:test11} show two examples of the algorithm, along with corresponding snapshots of a subset of the disks.

Figure~\ref{fig:CollStats} shows the mean number of collisions experienced by the disks per wave period,
and the mean magnitude of the accelerations caused by the collisions.
Accelerations occasionally exceeded $2g$ in two tests, which saturated the signals.
The mean accelerations calculated thus slightly underestimate the true mean.

Collisions occurred approximately once per wave period for the test using wave period $0.65$\,s and amplitude 10\,mm.
Surge amplitudes were not large enough to cause collisions in this test.
Collisions were, rather, caused by out of phase pitch motions of adjacent disks in the same column of the array.
However, the collisions were very weak, and did not force collided disks to separate. 
Thus, the initial configuration of the array was approximately maintained during the test, 
as shown in the left-hand panel of figure~\ref{fig:test06}.

Collisions occurred more than three times per period for the tests using a 2\,s wave period.
Surge amplitudes of the disks were largest in these tests.
However, adjacent disks surged almost exactly in phase with one another.
Small differences in the phase of surge motions caused disks to collide, but, typically, weakly.

Collisions occurred approximately twice per wave period for mid-range periods.
In the mid-range regime the collisions were at their strongest, 
as surge amplitudes and phase differences were large simultaneously.  
Strong collisions caused collided disks to rebound and collide with the opposing disk.
Hence, two collisions tended to occur in quick succession.
This behaviour can be inferred from the right-hand panel of figure~\ref{fig:test11}.

The strength of collisions in the tests using wave period 1.4\,s are strongly amplitude dependent.
Increasing the amplitude from 20\,mm to 40\,mm results in
mean accelerations increasing by at least a factor 2.7. 
(The true factor is unknown as the accelerations saturated the signal for the large-amplitude test.)
Weak dependence of energy transmission on wave amplitude for tests using wave period 1.4\,s
is therefore attributed to dependence collision strength on wave amplitude.

It is noted here that collisions can turn into rafting events when (i) relative motions of the disks allow for non-planar overlapping of edges,
and (ii) the collisions are not so strong as to cause disks to separate due to a glancing collision.
Rafting events occurred in three tests, with respective wave periods and amplitudes 0.8\,s and 40\,mm, 0.95\,s and 30\,mm, and 1.1\,s and 40\,mm.
Rafting events tend to develop in chains.
Chains of up to four disks were created in the tests conducted.

Figure~\ref{fig:Rafts} shows an example of a rafting event.
The disk in the middle row and left-of-centre column is rafted for almost the full duration of the steady-state window, starting at 76\,s and lasting beyond 96\,s.
The corresponding disk in the right-of-centre column is not rafted during the steady-state window.
The low-pass filtered acceleration time series for the two disks indicates that rafting suppresses the motion of the disk.
Rafting also reduces the number of collisions experienced by the disk.
The rafted disk experienced 22 collisions in the steady-state window, compared to 42 collisions experienced by the disk that did not raft.
However, no clear relationship is found here between rafting events and wave energy transmission.


\section{Summary and discussion}


Results of the first experimental campaign to investigate transmission of regular waves by arrays of thin floating disks in a wave basin have been reported.
The disks were wooden, with radius 0.495\,m and thickness 33\,mm.
A low-concentration array, consisting of forty disks, and a high-concentration array, consisting of eighty disks, were considered.
Wave periods in the range 0.65\,s to 2\,s were used. 
Wave amplitudes of 10\,mm, 15\,mm and 20\,mm were used for periods 0.65\,s to 0.8\,s, 0.95\,s and 1.1\,s to 2\,s, respectively. 
Larger wave amplitudes were also used for selected wave periods.
The transmitted wave field was recorded with a group of ten wave probes.
The key findings of the campaign are as follows.
\begin{itemize}
\item
Wave energy transmission increases monotonically, as wave period increases.
Wave transmission ranged from almost zero to full for the wave periods considered. 
\item
The high-concentration array transmits less energy than the low-concentration array.
However, wave energy transmission is not a simple function of concentration for mid-range wave periods, approximately 1.1\,s to 1.85\,s.
The lack of a simple relationship between transmission and concentration is attributed to additional wave energy attenuation due to collisions between disks
for the tests using a high-concentration array.
\item
The proportion of wave energy transmitted is strongly dependent on wave amplitude for periods less than 1\,s.
Doubling wave amplitudes, in this regime, caused transmitted energies to decrease by factors approximately 0.6 and 0.35 for the low and high-concentration arrays, respectively.  
Reduced wave energy transmission for large-amplitude waves and small wave periods
is attributed to additional energy attenuation due to the onset of wave overwash of the disks.
\item
The proportion of wave energy transmitted is weakly dependent on wave amplitude for mid-range wave periods and the high-concentration array.
Doubling the wave amplitude for period 1.4\,s caused the proportion of transmitted energy to decrease by a factor of approximately 0.93.
Reduced wave energy transmission for the large-amplitude wave, in this case, is attributed to stronger collisions, and hence increased energy attenuation.
\end{itemize}

Predictions of the energy transmitted by the arrays given by theoretical models 
were compared to transmitted energies extracted from the experimental data.
The models used are based on linear, potential-flow theory to model the water, and
thin-plate theory to model the disks.
Wave energy attenuation is due to an accumulation of scattering events.

Two two-dimensional models were considered. 
Backscatter is neglected in one model. 
Multiple scatterings between disks are incorporated in the alternative model, 
and the phase of the waves between the disks is considered to be random.
Both models provide the same expression for the energy transmitted by the disks.
A three-dimensional model was also considered.
The model is a version of the Boltzmann model, and employs the single-scattering approximation.

The key findings of the comparison of the theoretical models and the experimental data are as follows.
\begin{itemize}
\item
The two- and three-dimensional models give almost identical predictions of the transmitted energy.
\item
Model predictions of the transmitted energy agree 
with the experimental data for almost all tests using the low-concentration array. 
The only exception is the large-amplitude test using wave period 0.95\,s, 
\ie the test in which wave overwash of the disks appears to attenuate energy.
\item
For the high-concentration array, the theoretical models agree with the experimental data only for tests 
using the largest wave period, $\tau=2$\,s, in which case all of the incident wave energy is transmitted,
and wave periods less than or equal to 0.95\,s.
For wave periods 1.1\,s to 1.85\,s, \ie the tests for which wave energy appears to be attenuated by collisions,
models significantly overestimate the wave energy transmitted.
The overestimation is up to a factor of approximately 1.5. 
\end{itemize}

Model-data agreement for the low-concentration array and small wave amplitudes implies that in this regime:
\begin{enumerate}
\item[(i)]
scattering is the dominant source of wave attenuation; and
\item[(ii)]
linear, potential-flow/thin-plate theory provides a valid model to determine wave transmission through an array of floating disks.
\end{enumerate} 
Moreover, model-data agreement indicates that multiple wave scatterings do not affect wave transmission.
However, the significance of multiple scattering on wave transmission cannot be dismissed with the present investigation, 
because (i) multiple scattering is included in one of the two-dimensional models; 
and (ii) no extension of this model to three-dimensions exists (to the authors' knowledge).

The pronounced loss of model-data agreement for large-amplitude waves and the high-concentration array 
indicates that other sources of wave energy attenuation must be considered in these regimes.
Arguments have been given here that wave overwash of the disks and disk collisions are the additional sources of wave energy attenuation.
Dedicated experimental and modelling campaigns are required to determine qualitative relationships between overwash and collisions, and wave attenuation.




\section*{Acknowledgements}

The Collaborative \& Innovative Technology Program in Exploration and Production of Hydrocarbons
funded the experimental campaign,
with major sponsors Total, Saipem and {Doris} Engineering. 
The authors thank Francois Petrie, Vincent Lafon, Thierry Rippol and Alexandre Cinello (Oceanide, La Seyne Sur Mer) 
for helping design and conduct the experimental campaign.
The authors also acknowledge participation of Dany Dumont (Universit\'{e} de Quebec \`{a} Rimouski) in the experimental campaign.
The authors are grateful to Fabien Montiel (University of Otago) for providing advice on data analysis.
LB acknowledges funding support from the Australian Research Council (DE130101571) and the Australian Antarctic Science Grant Program (Project 4123). 
TW acknowledges funding support from the Norwegian Research Council, and Total E{\&}P (\textsc{wifar} project) and the U.S. Office of Naval Research (Award no: N62909-14-1-N010).


\end{document}